\documentclass[runningheads]{llncs}
\usepackage[T1]{fontenc}
\usepackage{graphicx}
\usepackage{amsmath}
\usepackage{amsfonts}
\usepackage{hyperref}
\usepackage{booktabs}
\begin{document}
\title{Alzheimer's disease detection based on large language model prompt 
engineering}
\author{Tian Zheng\inst{1,2} \and
Xurong Xie\inst{1\thanks{Corresponding Author}} \and
Xiaolan Peng\inst{1}\and
Hui Chen\inst{1} \and
Feng Tian\inst{1}}
\authorrunning{T. Zheng et al.}
\institute{Institute of Software, Chinese Academy of Sciences, Beijing 100190, China \\
\email{\{xurong,xiaolan,chenhui,tianfeng\}}@iscas.ac.cn\and
University of Chinese Academy of Sciences, Beijing 100049, China
\email{zhengtian24@mails.ucas.ac.cn}\\
}
\maketitle              % typeset the header of the contribution
\begin{abstract}
In light of the growing proportion of older individuals in our society, the timely diagnosis of Alzheimer's disease has become a crucial aspect of healthcare. In this paper, we propose a non-invasive and cost-effective detection method based on speech technology. The method employs a pre-trained language model in conjunction with techniques such as prompt fine-tuning and conditional learning, thereby enhancing the accuracy and efficiency of the detection process. To address the issue of limited computational resources, this study employs the efficient LORA fine-tuning method to construct the classification model. Following multiple rounds of training and rigorous 10-fold cross-validation, the prompt fine-tuning strategy based on the LLAMA2 model demonstrated an accuracy of 81.31\%, representing a 4.46\% improvement over the control group employing the BERT model. This study offers a novel technical approach for the early diagnosis of Alzheimer's disease and provides valuable insights into model optimization and resource utilization under similar conditions. It is anticipated that this method will prove beneficial in clinical practice and applied research, facilitating more accurate and efficient screening and diagnosis of Alzheimer's disease.

\keywords{Prompt learning ;  Large language model ;  Alzheimer's disease.}
\end{abstract}
\section{Introduction}
In recent years, the ageing of the world's population has become one of the new features of demographic change. With advances in medical care, the average life expectancy of human beings has lengthened, country-specific differences in population health have become more pronounced, and the prevalence of chronic diseases has risen globally. China is a large country with a large population and a seriously aging population. As the severity of aging increases, a variety of geriatric diseases follow, including Alzheimer's disease (AD), Parkinson's disease type of neurodegenerative diseases, and some statistics \cite{1} show that there are about 15 million dementia patients and more than 3 million Parkinson's patients among the elderly aged sixty years and above in China.Considering the global imbalance of healthcare resources, it is more difficult to diagnose AD by clinical visits, extensive neuropsychological testing or invasive means, but digital health tools can ideally address such issues. Completing a digital assessment at home via a smartphone or tablet would greatly increase the accessibility of Alzheimer's screening. 

The current speech-language data used for AD research suffers from high difficulty in data collection, limited number of samples, insufficient data diversity, and data quality problems. And the big language model can effectively deal with the above challenges. The large amount of pre-training data can compensate for the sparsity of data; the robust deep feature extraction of the large model can solve the problem of large individual differences in data\cite{13}; and compared with ordinary neural network models, the large language model can extract multi-level linguistic features that can help to comprehensively assess the patient's linguistic ability and capture potential AD symptoms.

In summary, this paper proposes the use of a large language model (LLAMA2) combined with a prompt engineering approach for AD detection method. The specific research methods are 1) using Prompt Learning combined with LORA to fine-tune the model; 2) using Prompt Tuning method to fine-tune the promptd utterances; and 3) using Conditional Learning to fine-tune the model. Experiments were conducted on the ADReSS2020 dataset, in which the method using Prompt Learning combined with LORA fine-tuning model achieved a classification accuracy of 81.31\%, which is higher than the BERT-based Prompt Learning method (76.85\%).
\section{Related work}
The extant literature on the detection of AD using speech-language data can be classified into three principal categories. The first category comprises studies that employ machine learning classification using manually extracted features. The second category encompasses studies that utilize automatic extraction of embedded features and classification using deep neural networks, among other techniques. The third category includes studies that employ fine-tuning and classification using pre-trained large language models.

Manually extracted features can be classified into two main categories: language-related features and acoustic-related features. Linguistic features encompass syntax, semantics, fluency/pause, word frequency, discourse continuity, and readability. Acoustic features, while not exhaustive, include spectral, rhythmic, jittery, and tonal qualities, as well as paralinguistic features such as rate of speech, phoneme accuracy, and so forth. In a previous study, the authors \cite{2}achieved a classification accuracy of 73.9\% using a simple Bayesian classifier after extracting the features contained in the eGeMAPS feature set. The eGeMAPS feature set\cite{3} is based on the OpenSMILE tool and the associated feature design and consists of 88 acoustic features, the values of which can be computed by analyzing the recording clips. These features encompass a range of characteristics, including frequency-dependent attributes such as pitch, jitter, and resonance peaks, as well as energy-dependent functions like shimmer, loudness, and the harmonic noise ratio. Additionally, they include spectral parameters like the alpha ratio and the Hammarberg ratio, harmonics, and ratios related to six temporal features. 

Furthermore, neural extraction is a prevalent methodology for implicit embedding characterization. The Distillable BERT model, as described in the literature\cite{5}, is a lightweight version based on the Bidirectional Encoder Representations from Transformers (BERT)\cite{14} model. The complexity of the BERT model is reduced through the process of model distillation, thereby enhancing the inference speed and resource efficiency of the model. The data set is then fed into the back-end classifier, which achieves an 88\% classification accuracy. In the literature\cite{6} , a Google VGGish model, pre-trained through Google's AudioSet, is employed to transform audio input features into linguistically and semantically meaningful high-level 128-dimensional embeddings.  Wav2Vec 2.0, as detailed in literature\cite{8}\cite{9}, is a self-supervised learning framework for speech recognition. It learns high-quality speech representations directly from raw audio waveforms through self-supervised learning, offering the advantages of lower data requirements and multilingual support.

The advent of large language models that have been pre-trained on vast quantities of data has given rise to novel approaches for the detection of AD. Pre-trained language models can be classified into two main categories: masked language models and autoregressive generative language models. The most prevalent masked language model is BERT and its numerous variants, including RoBERTa and Sentence BERT. Supervised model fine-tuning can be performed directly using labeled text, or the classification task can be converted into a prediction token using methods such as prompt. Study \cite{10}demonstrated the feasibility of this detection method by using spontaneous speech for the first time to classify text embeddings extracted from a large amount of pre-trained semantic knowledge on GPT-3. Nevertheless, the considerable number of parameters inherent to these large language models presents a challenge. For instance, GPT-3 has 175 billion parameters. 
\section{Data set and experimental methods}
In order to solve the problem of high computational cost, this experiment adopts two solutions: 1) using a smaller scale model LLAMA2-7b; and 2) reducing the scale of the parameters to be computed by using a fine-tuning method such as LORA during the fine-tuning.
\subsection{Dataset}
The dataset  is the ADReSS (Alzheimer's Dementia Recognition through Spontaneous Speech) Challenge 2020 dataset. The dataset comprises transcribed text in CHAT (Codes for the Human Analysis of Transcripts) format. In particular, the dataset comprises 108 data sets, each of which contains a description of an image (Cookietheft) provided by the subject and the corresponding label (Health/AD). The data are distributed in a quantitative manner, with the healthy and AD groups each comprising 54 of the data set. Additionally, the language utilized in the data set is English. In order to effectively utilize this data for the training of large-scale language models such as LLAMA2, a series of preprocessing steps are required. Initially, the text data must be cleaned and normalized in order to ensure data quality and consistency. Subsequently, text processing operations, such as word splitting and the removal of labeled words, are necessary in order to provide a clean and uniform input for the model. 
\subsection{BERT-based Prompt Learning}
The present study employs BERT-based experiments as a baseline for comparison. The methodology employed involved fixing the prompt template and adjusting the model parameters during the training phase. The pre-training models selected are BERT and RoBERTa. Robustly Optimized BERT Approach (RoBERTa) \cite{11}is an improvement to BERT proposed by the Facebook AI Research team. Task reconstruction is achieved by reconstructing AD and non-AD classification tasks as labeled words in filled prompted phrases. The implementation is divided into the following steps:

(1) Prompt phrase design: manually design the prompt phrases and select appropriate labeling words. “Diagnosis is <MASK>.” is the main template used in this paper, in which “dementia” and “healthy” are used as label words for AD and non-AD respectively.

(2) Combination of text and prompt phrases: The speech transcription text is combined with prompt phrases and input into PLM for prediction. The logits of the corresponding tagged words are generated from the prediction results, and the probability distribution of the tagged words is calculated by Softmax layer and optimized based on the binary cross-entropy loss.
\subsection{LLAMA2-based Prompt Learning}
\begin{figure}
\includegraphics[width=\textwidth]{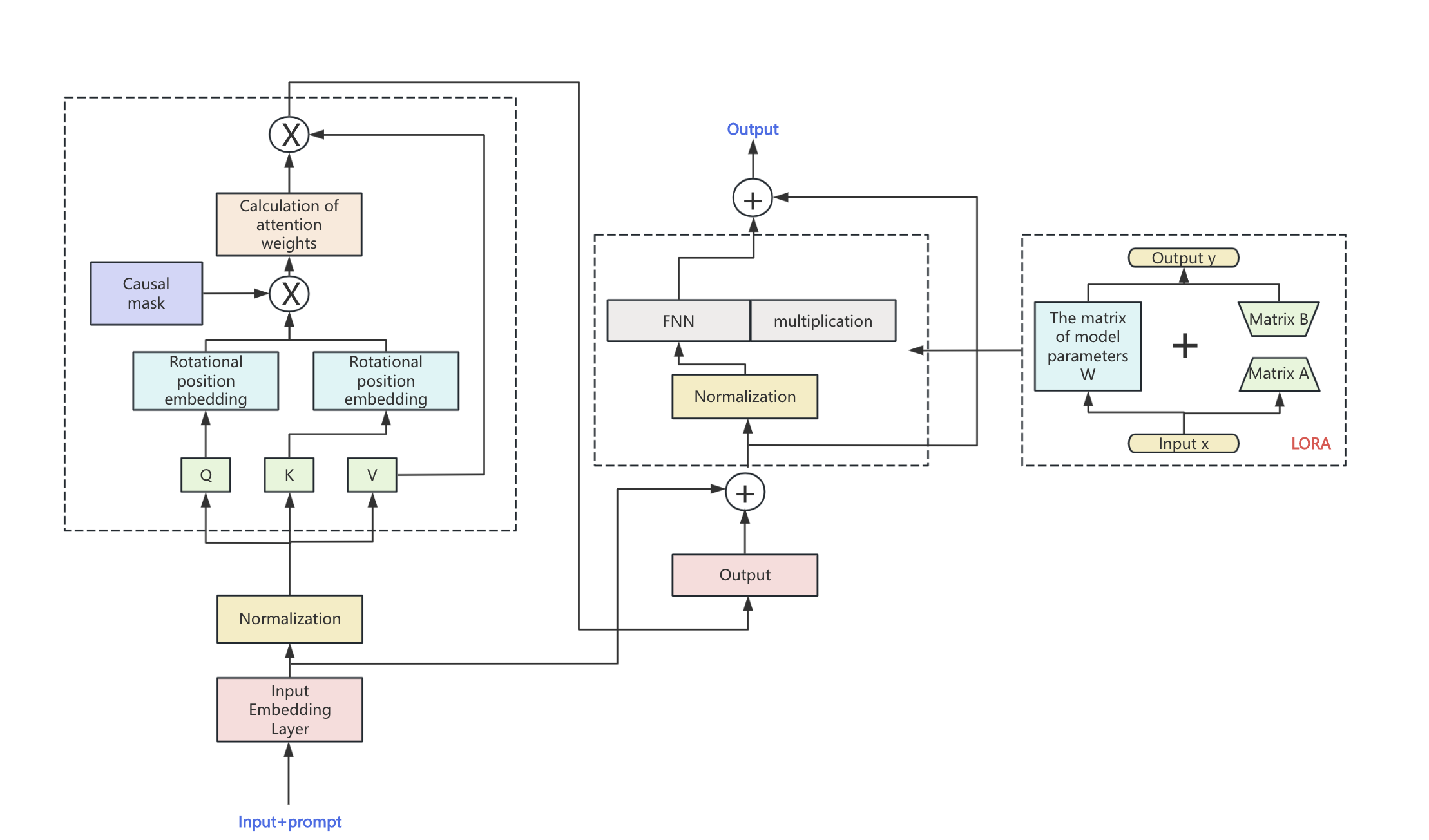}
\caption{Schematic diagram of the fine-tuning method based on the LLAMA2 model using prompt in combination with LORA.} \label{fig1}
\end{figure}

The experiments employed the Prompt Engineering with Frozen Transformers (PeFT) methodology and quantization techniques. The pre-training models were selected from the "Large Language Model Meta AI version 2," which was released by Meta.  The LLAMA2 model is based on the decoder-only architecture of Transformer, which is a neural network structure designed for text generation tasks. The model's architectural configuration is illustrated in Fig.\ref{fig1}.

The LLAMA2 model's decoder-only architecture and autoregressive generation mechanism facilitate effective performance in complex text generation tasks. The model is capable of discerning the subtleties in the input data and generating text of a high quality that is consistent with the context. Furthermore, as the model is designed with a focus on the decoding aspect, it is capable of demonstrating enhanced flexibility and efficiency in text generation, thereby providing users with expedient and precise responses.

The reconstruction of the text categorization task is achieved through the generation of responses based on prompts, which is facilitated by the utilization of LLAMA2. The implementation is divided into the following steps:

(1) Prompt phrase design: The prompt template is manually designed, and suitable labeling words are selected. The main template used in this experiment is "Input: {input} Instruction: {instruction} Response:". The response template utilized in this experiment is as follows:

(2) The combination of text and prompt phrases: The text data should then be combined with the prompt phrases and input into the model for prediction.

(3) Model quantization: The model weights are loaded into memory in the form of 8-bit integers using the quantization technique provided by the BitsAndBytes library, thereby improving the running efficiency of the model.

(4) LoRA Adaptation: The model is fine-tuned by introducing low-rank matrices in the key parts of the model.

Low-Rank Adaptation of Large Language Models (LoRA) is an approach to reduce the computational resource requirements for fine-tuning large language models (LLMs). The core idea of LoRA is to insert a pair of special low-rank projection matrices in each layer of the pre-trained model, thus restricting the update space of model parameters to these low-rank matrices. Specifically, LoRA introduces two small projection matrices, $\mathbf{A}$ and $\mathbf{B}$. Matrix $\mathbf{A}$ is of the form $\mathbf{A} \in \mathbb{R}^{r \times m}$, where $m$ and $n$ are the dimensions of the input and output layers, respectively. Matrix $\mathbf{B}$ is of the form $\mathbf{B} \in \mathbb{R}^{n \times r}$. The rank value $r$ is much smaller than $m$ and $n$. The projection matrices $\mathbf{A}$ and $\mathbf{B}$ represent the dimensions of the input and output layers, respectively.

The prompt-based fine-tuning method can effectively solve the problem of inconsistency between the loss function of the pre-trained model and the downstream task goal without changing the main structure of the pre-trained model.The overall schematic diagram is shown in Fig.~\ref{fig1}.

\subsection{LLAMA2-based Prompt Tuning}
The fundamental premise of Prompt-Tuning is that by fixing the parameters of the pre-trained model and adjusting only the parameters of the prompt statement component, the model is compelled to generate outputs that are more aligned with the task requirements specified by the prompt statements. 

Specifically, Prompt-Tuning comprises the following steps: firstly, an initial prompt statement is designed, based on the task requirements for cognitive impairment detection; secondly, an initial prompt statement template is created. To illustrate, the prompt utterance "Is there a speech disorder in this text?" can be designed to guide the model in recognizing the speech abnormality in the dialogue.

Once the model has been loaded, the next stage is to fine-tune the prompt parameters. During the training process, the parameters in the prompt statements are adjusted by the back-propagation algorithm, and the performance of the model on the validation set is optimized by means of cross-entropy. In contrast to traditional hard prompts, which are typically fixed text fragments, soft prompts are learnable vectors. These vectors are incorporated into the input embedding of the model and optimized during training through techniques such as gradient descent. In general, soft prompts can be adapted to different task requirements in a flexible manner, thereby enhancing the model's generalization ability and ultimately leading to improved results.
\subsection{LLAMA2-based Conditional Learning}
In previous experiments, labels were predicted based on input text. The text is about the image description in the dataset. The experimental method of conditional learning is to use an autoregressive model to compute the probability of a text given a label and then compute the probability of the same text under two different labels. The text with the higher probability is the predicted text. During training, the label is linked to the text and given to the autoregressive model to predict the next token.

The reconstruction of the text categorization task is reconstructed by computing the generation loss and probability. This is done by using a template with the label and text, with the label before or after the text. The text constitutes a description of the image. The label may be either "The following passage has a speech disorder" or "There is no speech disorder in the following passage."
\section{Experiments}
The experimental environment consists of the following configurations: the GPU model used is NVIDIA A40, with a total of five GPUs; the Python version is 3.7.0; the CUDA version is 11.6.0; and the server operating system is Linux.
\subsection{BERT-based Prompt Learning}
The fine-tuning process was implemented using the OpenPrompt framework, which is based on Pytorch for prompt fine-tuning. The BERT and RoBERTa models (bertbase-uncased and roberta-base) were selected for use, with their standard splitters employed in each case. The maximum length of the input text is set to 512. The hyperparameters are optimized on the cross-validation (CV) set through a greedy search. The hyperparameters that were prompted for fine-tuning included the learning rate, batch size, and AdamW optimizer. A weight decay of 0.01 should be applied to the LayerNorm module. Ten rounds of prompt fine-tuning are to be performed, with the 10-fold CV average accuracy serving as the scheduling criterion. The outputs of the final three epochs of fine-tuning are employed in a majority voting scheme to mitigate the risk of overfitting and to attenuate performance fluctuations.

\begin{figure}
\includegraphics[width=\textwidth]{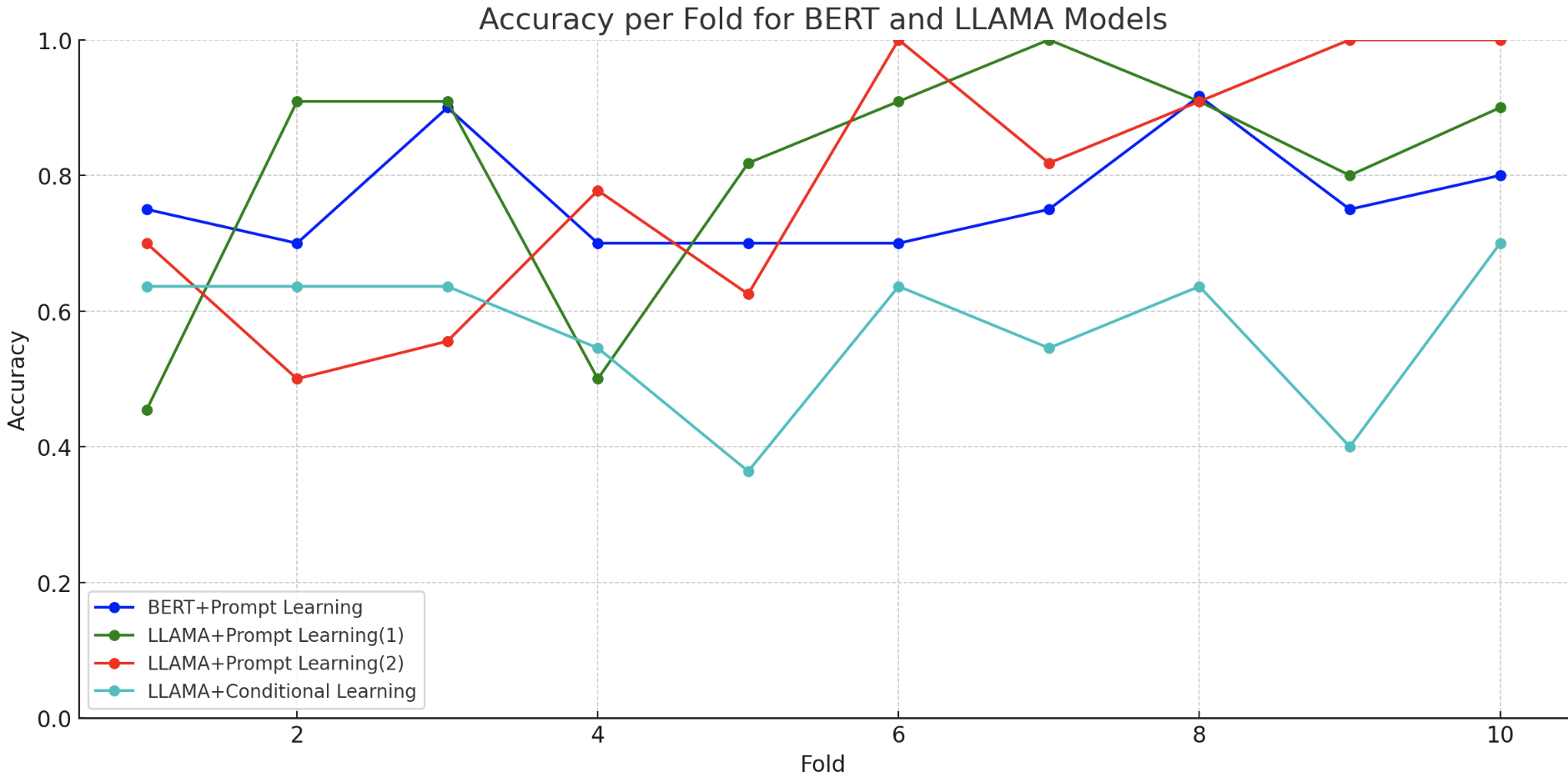}
\caption{10-fold cross-validation accuracy results.} \label{fig2}
\end{figure}

The prompted fine-tuning performance is as follows: The performance of the prompted fine-tuning of the PLM on the training data with 10-fold cross-validation is illustrated in Table~\ref{tab2}.The accuracy of each fold in the 10-fold cross-validation is shown in the Fig.~\ref{fig2}.
\begin{table}
\caption{Experimental results.}
\label{tab2}
\begin{tabular}{|l|l|l|l|l|l|l|l|}
\hline
Model & Approach & Accuracy & Precision & Recall & F1-score & Acc Std Dev& F1 Std Dev\\
\hline
BERT & Prompt Learning & 0.7685 & 0.8372 & 0.6667 & 0.7423 & 0.0816 & 0.1214\\
LLAMA2 & Prompt Learning(1) & \textbf{0.8131} & 0.8269 & 0.7963 & \textbf{0.8113} & 0.1844 & 0.1380\\
LLAMA2 & Prompt Learning(2) & 0.7938 & 0.7049 & 0.9556 & 0.8113 & 0.1890 & 0.1492\\
LLAMA2 & Conditional Learning & 0.5741 & 0.5769 & 0.5556 & 0.5660 & 0.1112 & 0.2271\\
\hline
\end{tabular}
\end{table}
\subsection{LLAMA2-based Prompt Learning}
The fine-tuning process was implemented using the PeFT framework. The LLAMA2-7b model was selected for use, and its standard splitter was employed. The maximum length of the input text is set to 512. The hyperparameters are optimized on the training set through cross-validation. The fine-tuned hyperparameters include the learning rate, microbatch size, gradient accumulation step, and AdamW optimizer. The model was subjected to multiple rounds of fine-tuning, with the 10-fold cross-validation average accuracy serving as the evaluation criterion.

1) Utilize the prompt "Is there a speech disorder to the above text?" and categorize the response as "No speech disorder/Presence of speech disorder."

2) Employ the prompt "Is there a dementia to the above text?" and categorize the response as "Healthy/Dementia."

The final 10-fold cross-validation performance obtained using the optimal model saved during training is presented in Table\ref{tab2}.
\subsection{LLAMA2-based Prompt Tuning}
The configuration of the tokenizer, the preprocessing of data, the fine-tuning of hyperparameters, and the evaluation of results are essentially the same as in the Prompt Learning section.

Due to the flexibility of the generative model, the effect of only training hints is very unsatisfactory. After many rounds of fine-tuning, the loss of cross-entropy is always greater than 1, and the results of many experiments do not exceed 50\% classification accuracy. Further experimental tuning is needed.
\subsection{LLAMA2-based Conditional Learning}
The configuration of the tokenizer, the preprocessing of data, the fine-tuning of hyperparameters, and the evaluation of results are essentially the same as in the Prompt Learning section.The same LORA fine-tuning method was used.

During the experiment, several rounds of fine-tuning verification are attempted, which may be limited by the computational resources, the results of several rounds are not very good, and the performance is shown in Table\ref{tab2}.
\section{Discussion}
In terms of average accuracy, the LLAMA2-based model (81.31\%) slightly outperformed the BERT-based method (76.85\%), suggesting that the LLAMA2 model performed better on the cognitive impairment classification task.

Although both methods used prompt fine-tuning, they differed in terms of model architecture, optimization techniques, and performance.The LLAMA2 model performed slightly better in the experiments, possibly because the input text was long descriptions, and the autoregressive nature of the generative model allowed for a more comprehensive understanding of the text.The seq2seq model was able to deal with longer contextual information, whereas the BERT model only needed to predict the words, and the generative model was can generate complete sequences.In addition, generative models are more flexible in adapting to new tasks and data distributions, especially when dealing with rare or new categories. In contrast, masking models require more fine-tuning to adapt to specific classification tasks.

The lower classification accuracy of the LLAMA2 model using the conditional learning approach may be due to label space constraints, the computation of generative probabilities, and the lack of label diversity in small datasets affecting the model's performance. The poor performance of cue fine-tuning in the AD classification task may be due to the complexity of the task making simple cue fine-tuning insufficient to capture and utilize complex patterns and features.

It is important to note that these advantages do not mean that the generative model outperforms the masked model in all scenarios. Masked models also perform well in many NLP tasks, especially in tasks that require contextual understanding and accurate prediction. Model selection should be based on specific task requirements, data characteristics, and performance goals.
\section{Conclusion}
 The objective of this paper is to examine the efficacy of a large-scale language model based on prompt engineering techniques for the efficient detection of cognitive impairment. The experiment initially selects prompt learning based on the BERT model as a performance baseline. Moreover, Prompt Learning and Prompt Tuning methods are innovatively explored in conjunction with the LLAMA2-7B model, with the objective of enhancing the model's capacity to identify cognitively impaired texts through the application of Prompt Engineering techniques. Additionally, experiments were conducted on conditional learning. To address the high computational demands of the large model, we employ a semi-precision quantization model and LoRA fine-tuning method, which effectively reduce the model's resource consumption. The experimental results demonstrate that the Prompt Learning method based on LLAMA2-7B attains an accuracy of 81.31\%, which is superior to the 76.85\% achieved by the control group. Nevertheless, there is considerable scope for improvement in the areas of Conditional Learning and Prompt Tuning, due to the limitations of their fine-tuning range and dataset size. To address this challenge, future work will focus on optimizing the Prompt Tuning and Conditional Learning strategies. Additionally, experiments will be extended to the Chinese dataset, with the aim of further improving the detection performance and practicality of the model, and providing strong technical support for the early diagnosis of cognitive impairment.

\begin{credits}
\subsubsection{\ackname}   The research work of this thesis was funded by the National Science and Technology Major Project on New Generation Artificial Intelligence (No. 2022ZD0118002) and the Basic Research Program of Institute of Software, Chinese Academy of Sciences (No. ISCAS-JCMS-202306).
\end{credits}
\bibliographystyle{splncs04}
\bibliography{arxiv}

\begin{thebibliography}{10}
\providecommand{\url}[1]{\texttt{#1}}
\providecommand{\urlprefix}{URL }
\providecommand{\doi}[1]{https://doi.org/#1}

\bibitem{10}
Agbavor, F., Liang, H.: Predicting dementia from spontaneous speech using large language models. PLOS digital health  \textbf{1}(12),  e0000168 (2022)

\bibitem{14}
Devlin, J., Chang, M.W., Lee, K., Toutanova, K.: Bert: Pre-training of deep bidirectional transformers for language understanding. arXiv preprint arXiv:1810.04805  (2018)

\bibitem{3}
Eyben, F., Scherer, K.R., Schuller, B.W., Sundberg, J., Andr{\'e}, E., Busso, C., Devillers, L.Y., Epps, J., Laukka, P., Narayanan, S.S., et~al.: The geneva minimalistic acoustic parameter set (gemaps) for voice research and affective computing. IEEE transactions on affective computing  \textbf{7}(2),  190--202 (2015)

\bibitem{8}
Gauder, M.L., Pepino, L.D., Ferrer, L., Riera, P.: Alzheimer disease recognition using speech-based embeddings from pre-trained models  (2021)

\bibitem{1}
Jia, L., Du, Y., Chu, L., Zhang, Z., Li, F., Lyu, D., Li, Y., Zhu, M., Jiao, H., Song, Y., et~al.: Prevalence, risk factors, and management of dementia and mild cognitive impairment in adults aged 60 years or older in china: a cross-sectional study. The Lancet public health  \textbf{5}(12),  e661--e671 (2020)

\bibitem{5}
Liu, N., Luo, K., Yuan, Z., Chen, Y.: A transfer learning method for detecting alzheimer's disease based on speech and natural language processing. Frontiers in Public Health  \textbf{10},  772592 (2022)

\bibitem{11}
Liu, Y., Ott, M., Goyal, N., Du, J., Joshi, M., Chen, D., Levy, O., Lewis, M., Zettlemoyer, L., Stoyanov, V.: Roberta: A robustly optimized bert pretraining approach. arXiv preprint arXiv:1907.11692  (2019)

\bibitem{2}
Luz, S., Haider, F., Fromm, D., Lazarou, I., Kompatsiaris, I., MacWhinney, B.: Multilingual alzheimer’s dementia recognition through spontaneous speech: a signal processing grand challenge. In: ICASSP 2023-2023 IEEE International Conference on Acoustics, Speech and Signal Processing (ICASSP). pp.~1--2. IEEE (2023)

\bibitem{6}
Mittal, A., Sahoo, S., Datar, A., Kadiwala, J., Shalu, H., Mathew, J.: Multi-modal detection of alzheimer's disease from speech and text. arXiv preprint arXiv:2012.00096  (2020)

\bibitem{9}
Pan, Y., Mirheidari, B., Harris, J.M., Thompson, J.C., Jones, M., Snowden, J.S., Blackburn, D., Christensen, H.: Using the outputs of different automatic speech recognition paradigms for acoustic-and bert-based alzheimer's dementia detection through spontaneous speech. In: Interspeech. pp. 3810--3814 (2021)

\bibitem{13}
Utama, P.A.: Robustness of pre-trained language models for natural language understanding  (2024)

\end{thebibliography}




% LNCS Homepage, \url{http://www.springer.com/lncs}, last accessed 2023/10/25
% \end{thebibliography}
\end{document}